\documentstyle[12pt]{article}
\evensidemargin
-0.5cm
\begin{document}
\begin{flushright}
INR-0904/95 \\
October  1995  \\
hep-ph/9510247
\end{flushright}
\begin{center}
{\large \bf BFKL QCD Pomeron in
High Energy Hadron Collisions \\
and Inclusive Dijet Production}
\footnote{To appear in {\it the Proceedings of the Workshop on Particle
Theory and Phenomenology}, the International Institute of
Theoretical and Applied Physics, Ames, Iowa, May 1995}
\\
\vspace{0.7cm}
{\large  Victor T. Kim${}^{\dagger}$
and Grigorii B. Pivovarov${}^{\ddagger}$ }\\
\vspace*{0.5cm}
{\em ${}^\dagger$ :
St.Petersburg Nuclear Physics Institute,
 188350 Gatchina,
Russia}\footnote{\em e-mail: $kim@fnpnpi.pnpi.spb.ru$}\\
\vspace*{0.3cm}
{\em ${}^\ddagger$ :
Institute for Nuclear
Research, 117312 Moscow, Russia}
\footnote{\em e-mail: $gbpivo@ms2.inr.ac.ru$}
\end{center}

\vspace*{0.5cm}
\begin{center}
{\large \bf
Abstract}
\end{center}

We calculate inclusive dijet production cross section in
high energy hadron collisions
within the BFKL resummation formalism
for the QCD Pomeron.
We take into account the Pomerons
which are adjacent to the hadrons. With these adjacent
Pomerons we  define a new
object --- the BFKL structure function of hadron --- which
enables one to calculate
the inclusive dijet production for any rapidity intervals.
We present predictions for the dijet  K-factor and
azimuthal angle decorrelation.
Estimations for some NLO BFKL corrections are also given.

\newpage

At present, much attention is being paid to the perturbative
QCD
Pomeron obtained by Balitsky, Fadin, Kuraev and Lipatov (BFKL)
\cite{Lip76}.
One of the reasons is that it relates hard processes
($ -t = Q^2 \gg {\Lambda^2_{QCD}}$) and
semi-hard ones ($s \gg -t =
Q^2 \gg {\Lambda^2_{QCD}}$):
It sums up leading energy logarithms of
perturbative QCD
into a singularity in the complex angular momentum
plane.
Several proposals to find direct manifestations of the BFKL
Pomeron
are available in the literature, see, e.g., Refs. [2-4], but
it is still difficult to get the necessary experimental data.

In this presentation we outline, within the BFKL approach, the
inclusive dijet cross
section in high energy hadron collisions
without any restrictions on untagging jets \cite{Kim95}.
Our goal is to push further towards the existing
experimental conditions the BFKL Pomeron predictions.

Removing the restriction on tagging jets to be most
forward/backward, which was imposed in the previous
studies, one should take into account additional
contributions to the cross section with jets more rapid than the
tagging ones. There are three such contributions:
two with a couple of Pomerons
(Figs. 1(b),1(c)) and one with three (Fig. 1(d)).
We will call the Pomerons developing between
colliding hadrons and their descendant jets the adjacent
Pomerons and the Pomeron developing between the tagging jets
the inner Pomeron. These additional contributions
contain extra power of $\alpha_S$ per extra Pomeron but hardly
could they be regarded as corrections since they
are also proportional to a kinematically
dependent factor which one can loosely treat as the number
of partons in the hadron moving faster
than the descendant tagging jet.

Mueller and Navelet result \cite{Mue87} for contribution
to the cross section
of Fig. 1(a) could be recast \cite{Kim95} as
\begin{eqnarray}
&&\frac{x_1x_2d\sigma_{\{P\}}}{dx_1dx_2d^{2}k_{1 \perp}d^{2}k_{2 \perp}}  =
\frac{\alpha_{S} C_A}{k^2_{1 \perp}}\frac{\alpha_{S} C_A}{k^2_{2 \perp}}
\times \nonumber \\
&&\sum_n\int d\nu
x_1 F_A(x_1,\mu^2_1)
\left[
\chi_{n,\nu}(k_{1 \perp})
e^{y\omega(n,\nu)}
\chi_{n,\nu}^*(k_{2 \perp})
\right]
x_2 F_B(x_2,\mu^2_2).
\label{cast}
\end{eqnarray}
where the subscript on $\sigma_{\{P\}}$ labels the contribution
to the cross section as a single inner Pomeron; $C_A=3$ is a color
group factor; $x_i$ are the longitudinal momentum fractions
of the tagging jets;
$k_{i\perp}$ are the transverse momenta;
$xF_{A,B}$ are the effective structure functions of colliding hadrons;
$y = \ln(x_1x_2s/k_{1\perp}k_{2\perp})$ is the relative rapidity
of tagging jets;
%\begin{equation}
$\chi_{n,\nu}(k_{\perp})=\frac{(k_{\perp}^2)^{-\frac{1}{2}+i\nu}
e^{in\phi}}{2\pi} $
%\label{eigenfunction}
%\end{equation}
are Lipatov's eigenfunctions and
$\omega(n,\nu) = \frac{2 \alpha_{S} C_A}{\pi}
\biggl[ \psi(1) - Re \, \psi \biggl( \frac{|n|+1}{2} +
i\nu \biggr) \biggr]$ are
Lipatov's eigenvalues. Here $\psi$ is the logarithmic derivative
of Euler Gamma-function.

   As we have shown \cite{Kim95}, subprocesses of Fig. 1(b)-1(d) with the
adjacent Pomerons contribute to the
 effective structure
functions, i.e., one can account for them by just adding some
``radiation corrections''  to the structure functions
of Eq.(\ref{cast}):
\begin{eqnarray}
x F_A(x_1,\mu^2_1) & \Rightarrow &x  \Phi_{A}(x_1,\mu^2_1,n,\nu,k_{1\perp})
 \equiv
x F_A(x_1,\mu^2_1) + x D_{A}(x_1,\mu^2_1,n,\nu,k_{1\perp}), \nonumber \\
x F_B(x_2,\mu^2_2)&\Rightarrow&
x \Phi_{B}^{\ast}(x_2,\mu^2_2,n,\nu,k_{2\perp})
\equiv
x F_B(x_2,\mu^2_2) + x D_{B}^{\ast}(x_2,\mu^2_2,n,\nu,k_{2\perp}),
\nonumber \\
&&
\label{subs}
\end{eqnarray}
where $x \Phi_{A,B}$ are the new structure functions that depend
on Lipatov's quantum numbers $(n,\nu)$ --- we call them BFKL
structure functions;
the complex conjugation on $\Phi_B$ could be understood
if one look at rhs of Eq.~(\ref{cast}) as a matrix element
of a $t$-channel evolution operator with the relative
rapidity, $y$, as an evolution parameter and $F_B$
as a final state; $(n,\nu)$ are then  ``good quantum numbers''
conserved under the evolution---this makes room for
$(n,\nu)$-dependence of the corrected structure functions.
We note also that the corrected structure functions
may depend on the transverse momenta of the tagging jets.
An explicit expression for the radiation correction  $D_{A,B}$
to the effective hadron structure functions see in Ref. \cite{Kim95}.

Eq.~(\ref{cast})  with the substitution (\ref{subs}), makes
possible to get updated predictions for the $K$-factor and the
azimuthal angle decorrelation of $x$-symmetric ($x_1=x_2$) dijets
on an effective relative rapidity
$y^{\ast}\equiv\ln({x_1x_2s}/{k_{\perp min}^2})$
(see Figs. 2,3, where the LO CTEQ3L structure functions \cite{Lai94}
have been used).

A look at the plots brings a conclusion that the adjacent Pomerons
may play a decisive role in the high energy hadron collisions.
We note also that one should not stick anymore to the large
dijet relative rapidity region in the BFKL Pomeron manifestations
hunting, since, from the one hand, we include the region of the
moderate rapidity intervals into our consideration and,
from the other hand, the resummation effects are quite pronounced
at the moderate rapidity region.

We present also in Figs. 2,3 estimations for NLO BFKL
effects using the results of Ref. \cite{Cor95}, where
conformal NLO contributions to the Lipatov's eigenvalues
were calculated. The estimations incorporate the NLO
conformal corrections to the Lipatov's eigenvalues
(see Fig. 4) and the NLO CTEQ3M structure functions~\cite{Lai94}.

We should note here that the extraction of data on
high-$k_{\perp}$ jets from the event samples in order to
compare them with the BFKL Pomeron predictions
should be different from the algorithms
directed to a comparison
with perturbative QCD predictions for the hard
processes. These algorithms, motivated by the strong
$k_{\perp}$-ordering of the hard QCD regime, employ
hardest-$k_{\perp}$ jet selection
(see, e.g., Ref. \cite{Alg94}). It
is doubtful that one
can reconcile these algorithms with the weak $k_{\perp}$-diffusion
and the strong rapidity ordering of the semi-hard QCD regime,
described by the BFKL resummation. We also note
that our predictions should not be compared with the
preliminary data \cite{Heu94} extracted by the most forward/backward
jet selection criterion. Obviously, one should include
for tagging all the registered pairs of jets
(not only the most forward--backward pair) to compare
with our predictions. In particular, to make a comparison
with Figs. 2,3, one should sum up all the registered
$x$-symmetric dijets ($x_1=x_2$)
with transverse momenta harder than $k_{\perp min}$.

 We thank E.A.Kuraev and L.N.Lipatov for stimulating discussions.
 We are grateful to A.J.Somme\-rer,
 J.P.Vary, and B.-L.Young for their kind hospitality at the
 IITAP, Ames, Iowa and support.
 V.T.K. is indebted to  S.Ahn, C.L.Kim, T.Lee, A.Petridis,
 J.Qiu, C.R.Schmidt, S.I.Troyan, and C.P.Yuan for helpful conversations.
 V.T.K. also thanks the Fermilab Theory Division for hospitality.
 G.B.P. wishes to thank F.Paccanoni for fruitful discussions and
 hospitality at the Padova University.

%%%\section*{References}

\section*{Figure Captions}

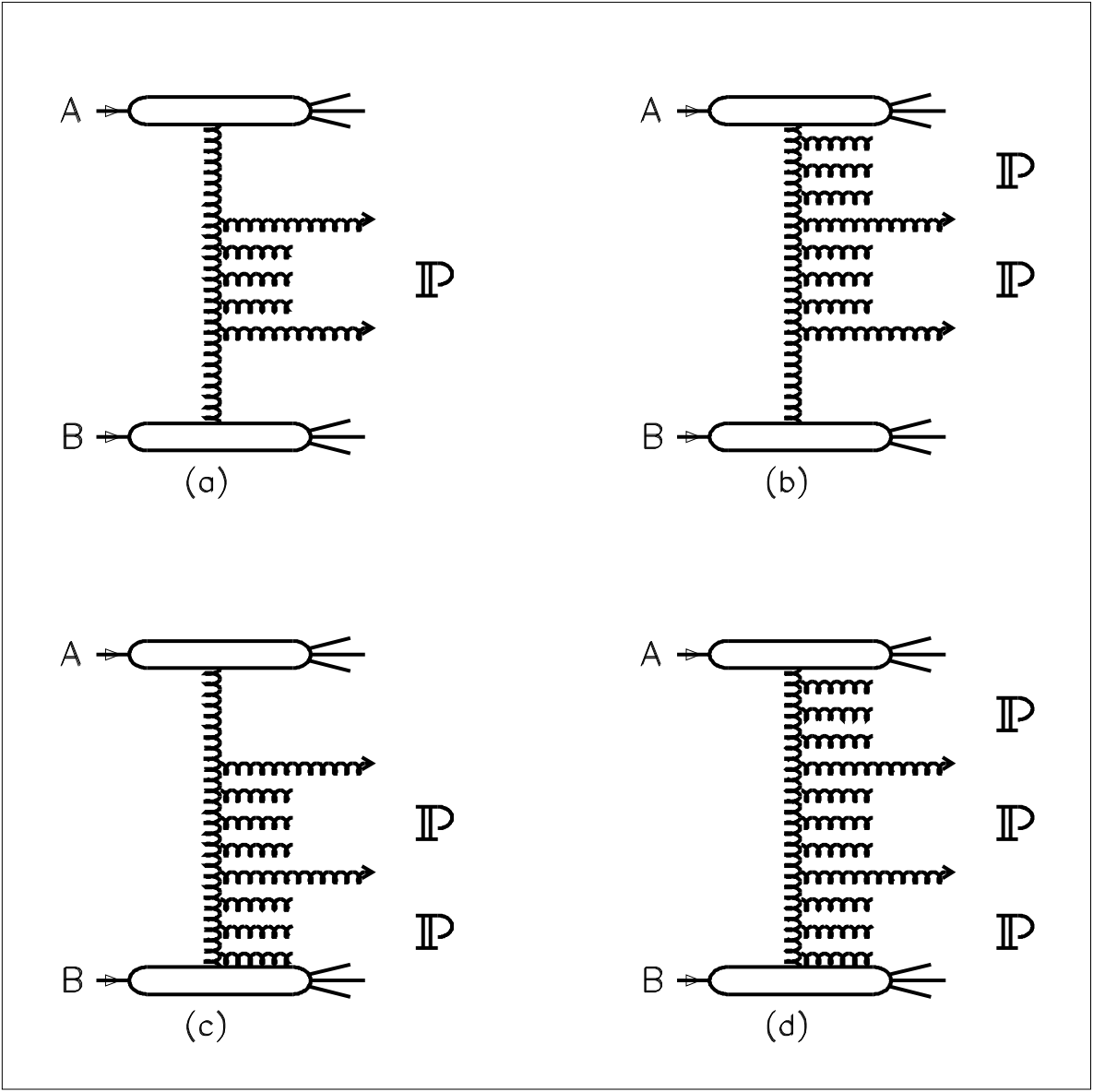
Fig. 1: Subprocesses for the dijet production in hadron collision.
\newline
Fig. 2: The $y^{\ast}$-dependence of the dijet K-factor.
\newline
Fig. 3: The $y^{\ast}$-dependence of the average azimuthal angle cosine
between the \\
\hspace*{1.1 cm} tagging jets.
\newline
Fig. 4: The Lipatov's eigenvalues at $n=0$.

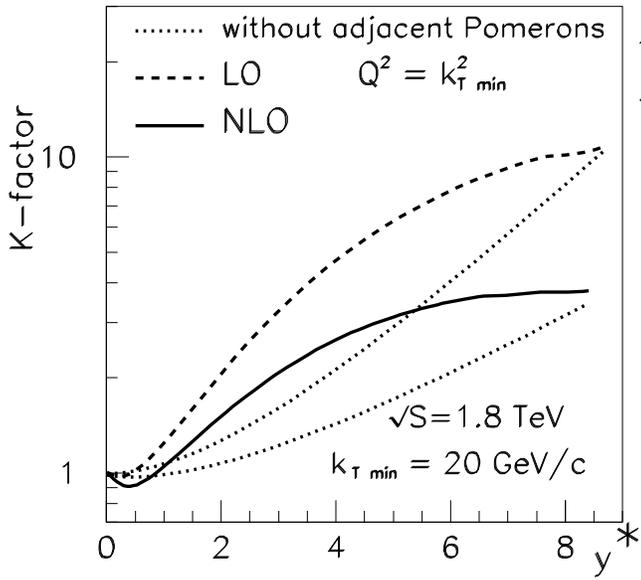
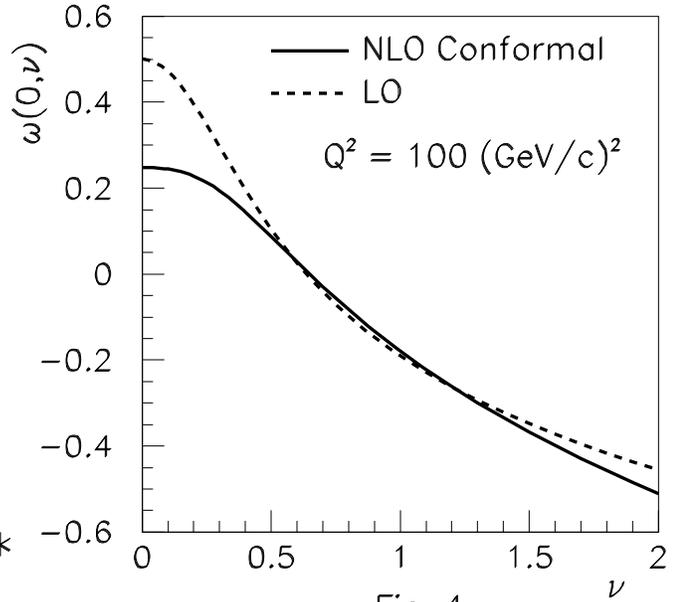
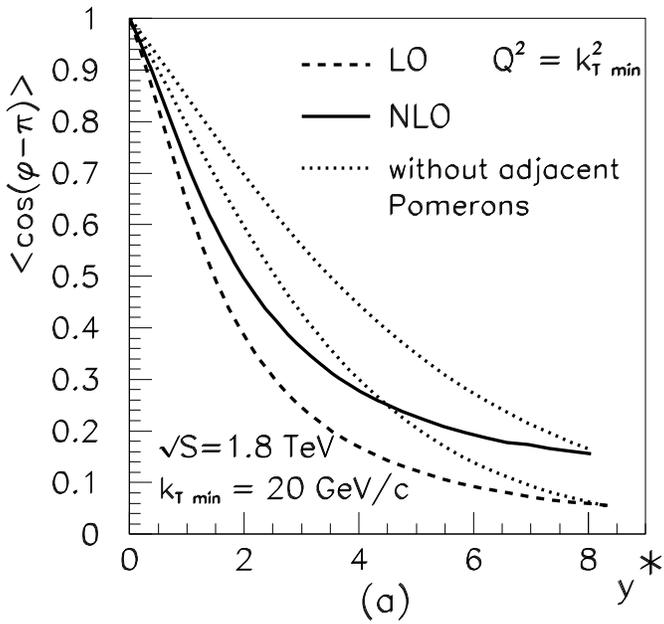
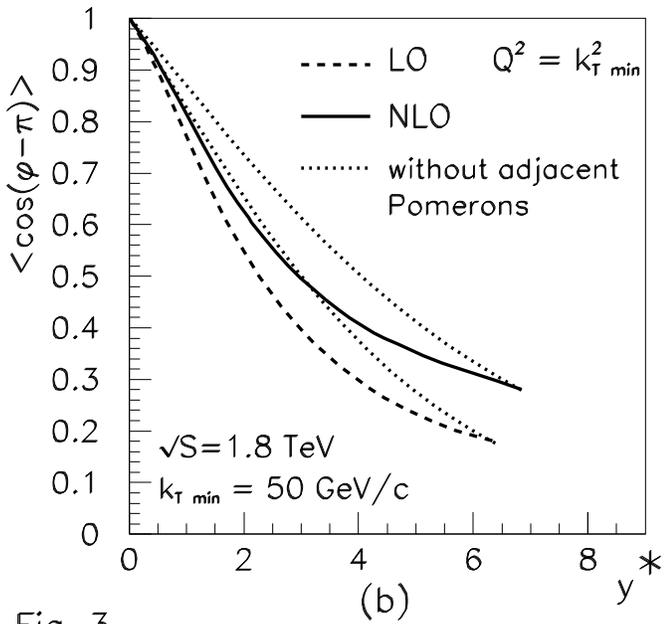
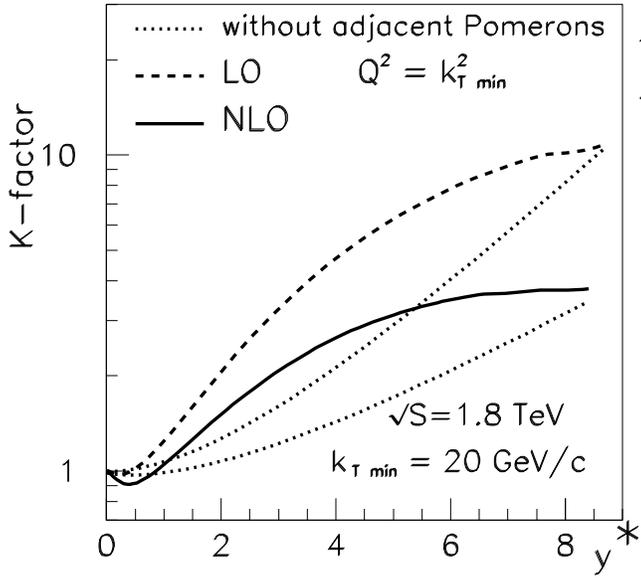
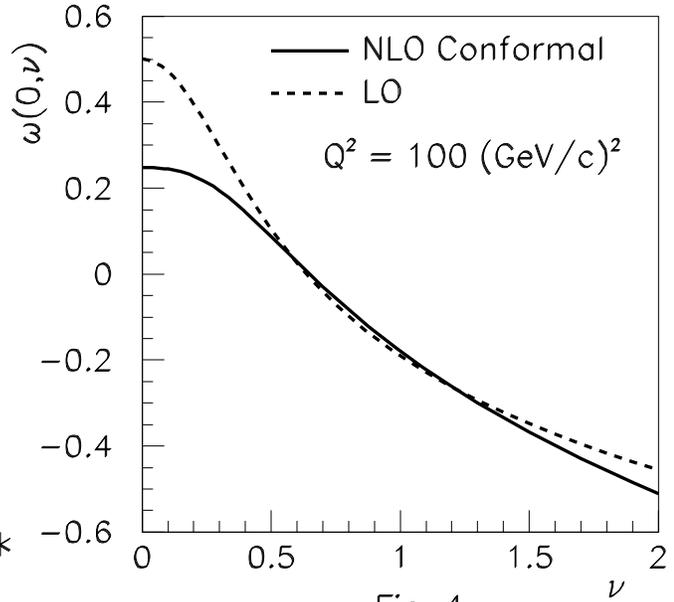
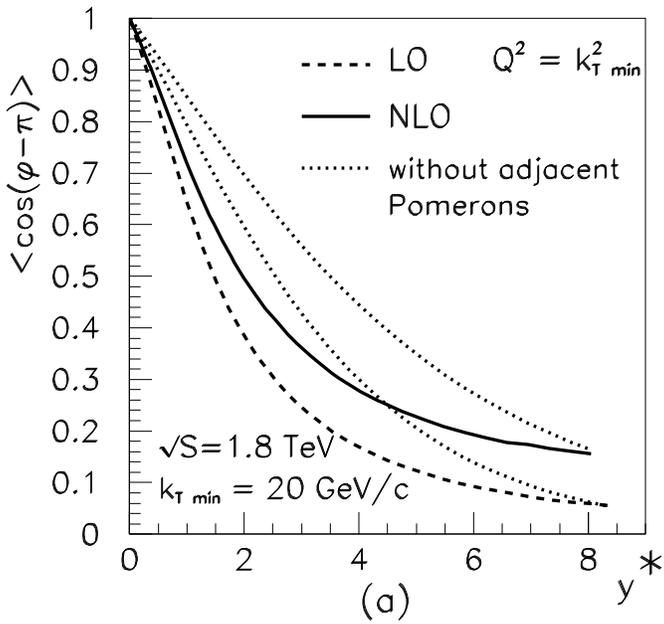
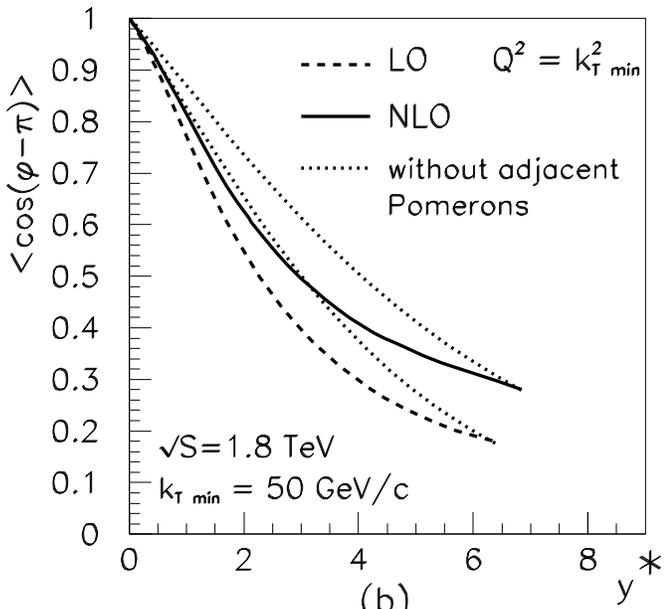
\end{document}